\documentclass[iop,onecolumn]{emulateapj}
\usepackage{natbib, aas_macros}
\shorttitle{Viability of MRI in CPD}
\shortauthors{Fujii et al.}

\begin{document}

\title{On the Viability of the Magnetorotational Instability 
in Circumplanetary Disks}

\author{Yuri I. Fujii\altaffilmark{1}, Satoshi Okuzumi\altaffilmark{12},
Takayuki Tanigawa\altaffilmark{3}, and Shu-ichiro Inutsuka\altaffilmark{1}}
\altaffiltext{1}{Department of Physics, Nagoya University, Furo-cho, Chikusa-ku,
Nagoya, Aichi 464-8602, Japan}
\altaffiltext{2}{Department of Earth and Planetary Sciences, Tokyo 
                  Institute of Technology, Meguro-ku, Tokyo, 152-8551, Japan}
\altaffiltext{3}{Institute of Low Temperature Science, 
                  Hokkaido University, Sapporo, 060-0819, Japan}
\email{yuri.f@nagoya-u.jp}

\begin{abstract}
{We examine whether the magnetorotational instability (MRI)
can serve as a mechanism of angular momentum transport
in circumplanetary disks.
For the MRI to operate the ionization degree must be sufficiently high
and the magnetic pressure must be sufficiently lower than the
gas pressure. We calculate the spatial distribution of the 
ionization degree and search for the MRI-active region where 
the two criteria are met.
We find that there can be thin active layers at the disk surface
depending on the model parameters, however,
we find hardly any region which can sustain well-developed MRI turbulence; 
when the magnetic field is enhanced by MRI turbulence at the 
disk surface layer, a magnetically dominated atmosphere encroaches on a lower
altitude and a region of well-developed MRI turbulence becomes smaller. 
We conclude that if there are no angular momentum transfer mechanisms 
other than MRI in gravitationally stable circumplanetary disks, 
gas is likely to pile up until disks become gravitationally unstable,
and massive disks may survive for a long time.}
\end{abstract}


\section{Introduction}
\label{sec:1}
Gas giant planets obtain mass from surrounding protoplanetary disks 
during their formation. When gas flows onto the planets, 
disks form around them as by-products. 
These gaseous disks are called circumplanetary disks.
The evolution of circumplanetary disks is important 
not only for the formation of gas giants but also for the 
formation of satellites because regular satellites are 
thought to form in circumplanetary disks.  
There have been several theoretical studies of satellite formation in disks
\citep[e.g.,][]{lun82, can02, can06, mos03a, mos03b, can09, est09, sas10, ogi12}. 
\citet{can02, can06} developed a so-called gas-starved disk 
model and successfully explained the total mass of the Galilean 
satellites and other satellite systems around giant planets
in our solar system. 
In their scenario, gas and solids are continuously 
supplied to circumplanetary disks and satellites form and migrate onto the central
planet repeatedly; the last generation of satellites remains
when the mass inflow terminates. 
\citet{mos03a, mos03b} developed a so-called solid enhanced 
minimum mass disk model and reproduced the Galilean satellites; 
the inner three satellites are formed in an inner massive disk and the outermost 
satellite is formed slowly in an outer extended low-density disk.  
A remaining problem is how to form circumplanetary 
disks accounting for their viscous evolution and mass infall 
from protoplanetary disks.
Recently, a new idea for satellite formation has been advocated by \cite{cri12}.
Using their gas-free tidal-spreading particle disk model, 
the mass distributions of most regular satellite systems 
in our solar system can be well reproduced.
However, it is difficult to formed Galilean satellites in this 
model, which would imply that
the satellites still need 
a gaseous circumplanetary disk for their formation. 

Hydrodynamic simulations of gas giant formation
\citep[e.g.,][]{lub99, tan02, ayl09a, ayl09b, mac06, mac08, mac10} 
have shown that circumplanetary disks form during the accretion 
phases of giant planets. 
Since most of these studies focused on gas giant formation, 
the detailed distribution of gas flow onto a circumplanetary disk was
not well investigated.
Recently, high-resolution 3D simulations by \citet{tan12} 
have demonstrated that gas flows onto circumplanetary disks from high altitudes, 
not from the disk mid-plane. 
This was the first detailed analysis of the flux of gas infall from
protoplanetary to circumplanetary disks.
This picture is consistent with radiative 
hydrodynamical simulations by \citet{kla06} 
and the very recent high-resolution global simulations by \citet{gre13} and \citet{ szu13}.

On the other hand, the driving mechanism of angular momentum transfer of gas 
in circumplanetary disks is not yet well understood. For accretion disks in general,
the most promising mechanism is believed to be the magnetic turbulence 
driven by the magnetorotational instability (MRI).
To be MRI-active, gas in a disk should be sufficiently ionized 
to couple with the magnetic field. The main ionization source 
in circumplanetary disks is galactic cosmic rays, and their 
attenuating length is about $100\ {\rm g\ cm^{-2}}$.
In some studies of MRI, a critical surface 
column density of $\Sigma_{\rm crit}\sim100\ {\rm g\ cm^{-2}}$ 
was adopted; a surface density below this value implies that 
the MRI can operate \citep{gam96a}. 
However, if we take into account the chemical 
reactions of charged particles, such as recombination or 
capture by dust grains, the critical value can be far smaller 
\citep*{san00, ilg06a, oku09, fuj11}.
In the context of circumplanetary disks, \citet{mar11a} and \citet{lub12} pointed
out the possibility of accretion outbursts induced by a combination
of MRI and gravitational instability (GI).

In this work, we develop a model for circumplanetary disks 
by calculating the surface density with the mass infall rate 
obtained by \citet{tan12}, and investigate whether the 
MRI is important in circumplanetary disks. Since \citet{tan12} have found
that the mass infall rate is proportional to the surface density of 
the parental protoplanetary disk, we can model the various evolution phases
of the disk. We introduce a gas depletion factor
and model the situation when a gas giant opens a gap, or gas in the 
protoplanetary disk is globally depleted.
For the evaluation of MRI activity, we use the Elsasser number.
The Elsasser number is proportional to the ionization degree, 
which we calculate for several conditions using 
the method developed by \citet*{fuj11}. 
For ionization sources, we take into account galactic cosmic rays, 
X-rays from the host star, and the decay of short-lived radionuclides.

In Section \ref{sec:2}, we explain how to estimate the border of
MRI-active/inactive regions in gaseous disks such as circumplanetary 
disks or protoplanetary disks. Our models of circumplanetary disks
are described in Section \ref{sec:CPDmodel}. In Section \ref{sec:MRIinCPD}, 
we investigate MRI activity in circumplanetary disks using 
our ionization degree calculation method described in 
Section \ref{sec:2}. A discussion of our results is given in Section
\ref{sec:5}, and we summarize this paper in Section \ref{sec:6}. 

\section{MRI and ionization degree}
\label{sec:2}
\subsection{Conditions for MRI growth}
\label{sec:2-1}
There are two criteria for the MRI to be active 
\citep{bal91, san99, oku11}.
First, the ionization degree of the disk gas should be high enough 
to couple to the magnetic field. We use the Elsasser number to investigate 
MRI activity. The Elsasser number is written as
\begin{eqnarray}
    \Lambda = \frac{v_{{\rm A}z}^2}{\eta \Omega_{\rm K}},
\label{411}
\end{eqnarray}
where $v_{{\rm A}z}$ is the $z$ component of the Alfv\'en 
velocity, $\eta$ is the magnetic diffusivity, 
and $\Omega_{\rm K}$ is the Keplerian frequency.
In order for the MRI to operate, $\Lambda$ must be larger than unity
\citep{san99}. The region where $\Lambda<1$ is a dead zone.
The magnetic diffusivity can be written 
as follows \citep{bla94}:
\begin{equation}
    \eta = 234\left( \frac{T}{1 {\rm K}} \right)^{1/2}x_{\rm e}^{-1}
    \ {\rm cm^2\ s^{-1}},
    \label{412}
\end{equation}
where $x_{\rm e}\equiv n_{\rm e}/n_{\rm n}$ (the ratio of the number 
densities of electrons and neutral gas molecules) is the ionization degree. 
Equations (\ref{411}) and (\ref{412}) show that $\Lambda$ is proportional
to the ionization degree, $x_{\rm e}$.
Here, we consider only Ohmic dissipation and neglect the ambipolar and Hall
diffusivities that mostly work to stabilize the MRI.
If we know the strength of the magnetic field and the temperature 
of the gas, we only need the ionization degree to estimate $\Lambda$
(note that ions and charged dust grains are much heavier than 
electrons and therefore their motion is negligible). 
In this work, we treat the magnetic field as a parameter 
and calculate the ionization degree of an isothermal disk. 
The ratio of gas pressure, $P_{{\rm gas}}$, and the $z$ component of
magnetic pressure, $P_{{\rm mag},z}$, represents the $z$ component 
of the plasma beta, which is
\begin{eqnarray}
    \beta_{\rm z} &\equiv& \frac{P_{\rm gas}}{P_{\rm mag,z}}
          = \frac{\rho_{\rm g}c_{\rm s}^2}{B_{\rm z}^2/8\pi}
          = \frac{2c_{\rm s}^2}{v_{{\rm A}z}^2},
    \label{528}
\end{eqnarray}
where $B_{\rm z}$ is the net vertical magnetic field, 
$v_{{\rm A}z} = B_{\rm z} / \sqrt{\mathstrut 4\pi \rho_{\rm g}}$
is the $z$ component of Alfv\'en velocity, $c_{\rm s}$ is the 
sound speed, which is assumed constant, and $\rho_{\rm g}$ 
is the gas density. Note that $\beta_{\rm z}$ is defined 
in terms of $net$ magnetic flux.
In this study, we assume that
the disk is vertically hydrostatic 
(see Equation (\ref{33}))
and that $B_{\rm z}$ is 
vertically constant. Thus, we write $\beta_{\rm z}$ as 
$\beta_0\exp(z^2/2H^2)$,
where $\beta_0$ is the mid-plane value of $\beta_{\rm z}$, 
$z$ is the height from the mid-plane, and $H$ is the scale
height of the disk. 
We assume $\beta_0$ is constant and choose $\beta_0=10^4$ and $10^5$,
which are optimistic values for MRI. 
The higher net field strength means that more magnetic flux
threads the disk.

Second, the wavelength of the most unstable mode,
$\lambda_{\rm MRI}=2\pi v_{\rm A}/\Omega_{\rm K}$, should be smaller
than the scale height of the disk. This corresponds to the condition for 
a weak magnetic field: if the magnetic field is too strong, magnetic tension
prevents disk gas from becoming turbulent.
We refer to the region with $\lambda_{\rm MRI}>H$ as the 
magnetically dominated atmosphere where the MRI is suppressed. Thus, from 
Equation (\ref{528}), the condition to be MRI-active is written as 
\begin{eqnarray}
    2\pi v_{\rm A}/\Omega_{\rm K} = \lambda_{\rm MRI}
    &<& H = c_{\rm s}/\Omega_{\rm K},
    \label{beta_condition}
\end{eqnarray}
or equivalently,
\begin{eqnarray}
    \beta_{\rm z} &>& 8\pi^2.
    \label{8pi}
\end{eqnarray}
%
When the MRI is driven, magnetic fields are amplified by turbulence 
and the magnetically dominated atmosphere encroaches on the active region. 
\citet{oku13a} quantitatively evaluated this 
using MHD simulations by \citet{oku11} and \citet{gre12}. 
At the border separating active from dead zones, 
$B_{\rm z}^2$ is amplified to be roughly 30 times larger than 
the original value. \citep[see Equation (37) of][]{oku13a}.
This implies that the real plasma beta at the border becomes $\beta_{\rm z}/30$. 
Thus, when the turbulence is well-developed, the criterion to have MRI
becomes $\beta_{\rm z}/30 > 8\pi^2$, or approximately,
\begin{eqnarray}
    \beta_{\rm z} &\gtrsim& 2000.
    \label{2000}
\end{eqnarray}
%
The lower limit of the magnetically dominated atmosphere is located between
1.8$H$ and 3.8$H$, depending on $\beta_0$ and the criteria of the second
condition (see Figs. \ref{fig:f_dp=1}, \ref{fig:f_dp=10^-3}, and 
\ref{fig:f_dp=10^-5}). We ignore the effect of mixing on chemistry 
because it is important only above about 3$H$, or even higher when there
are dust grains.
The onset of the MRI occurs only in the region where the two conditions
of Equation (\ref{411}) and (\ref{8pi}) are both met. The conditions to
have well-developed MRI turbulence are severer than those of just onset
of the MRI, which are Equation (\ref{411}) and (\ref{2000}).
\subsection{Calculation of ionization degree}
We calculate the ionization degree accounting for dust grains. 
We assume that there are plentiful metal ions so that molecular ions
transfer their charge to metal ions quickly.
We use the following rate equations derived by \citet*{fuj11} based on
\citet{opp74} and \citet{oku09}:
\begin{eqnarray}
    \frac{dn_{\rm M^+}}{dt} &=& \zeta n_{\rm n} - \alpha_{\rm M^+}n_{\rm M^+}n_{\rm e} 
       -\langle k_{\rm M^+ d}\rangle N_{\rm d} n_{\rm M^+},\\
    \label{261}
    \frac{dn_{\rm e}}{dt} &=& \zeta n_{\rm n} - \alpha_{\rm M^+}n_{\rm M^+}n_{\rm e} 
       -\langle k_{\rm e d }\rangle N_{\rm d} n_{\rm e},\\
    \label{262}
    \frac{d\langle Z \rangle}{dt}&=&
        \langle k_{\rm M^+ d} \rangle n_{\rm M^+} 
        -\langle k_{\rm e d} \rangle n_{\rm e}, \\
\label{263}
%
    \frac{d\langle \delta Z^2 \rangle}{dt}
    &=& \left( \langle k_{\rm M^+ d} \rangle 
           +2\langle  k_{\rm M^+ d} \delta Z \rangle \right)n_{\rm M^+}\nonumber\\
    &\ \ &  +\left( \langle k_{\rm e d} \rangle
            -2\langle k_{\rm e d} \delta Z \rangle \right)n_{\rm e}.
\label{264}
\end{eqnarray}
where $n_{\rm j}$ indicates the number density of each
particles ($\rm n$, neutral molecules; $\rm e$, electrons; 
$\rm M^+$, metal ions), $N_{\rm d}$, $\langle Z \rangle$, and
$\langle \delta Z^2 \rangle$ are the total number density, 
mean charge, and dispersion of the charge distribution
of dust grains, respectively,
$\zeta$ is the ionization rate, $\alpha_{\rm M^+}$ is the reaction rate 
of radiative recombination, and $\langle k_{\rm jd} \rangle$ is the 
capture rate onto a dust grain surface weighted by 
the number density of dust grains of charge $Z$.
We use the UMIST database (RATE`06) for
$\alpha_{\rm M^+}=2.80\times 10^{-12}
(T/300\ {\rm K})^{-0.86}\ \ \rm cm^3\ s^{-1}$,
where $T$ is temperature. 
%
In this work, we assume compact spherical dust grains with
density $\rho_{\rm grain}=3\ \rm g\ cm^{-3}$ and
radii $a=0.1$ and $10{\rm\ \mu m}$.
The mass of a grain is 
$m_{\rm grain} = (4\pi/3)\rho_{\rm grain}a^3$.
We define the dust-to-gas mass ratio as the ratio of 
spatial density of dust grains, $\rho_{\rm d}$, to that of gas, 
$\rho_{\rm g}$:
\begin{eqnarray}
    f_{\rm dg} \equiv \frac{\rho_{\rm d}}{\rho_{\rm g}},
\label{247}
\end{eqnarray}
and use $f_{\rm dg} = 10^{-2}$. We can write the number density 
of dust grains as $n_{\rm d}= \rho_{\rm d}/
m_{\rm grain}=f_{\rm dg}\rho_{\rm g}/m_{\rm grain}$.

\subsection{Ionization rate}
\begin{figure}[htpb]
    \epsscale{0.50}
    \plotone{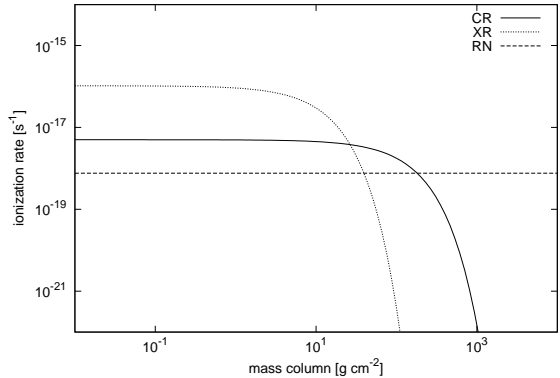}
         \caption{Ionization rates due to cosmic rays (solid line),
         X-rays (dotted line), and radionuclides (dashed line)
         as a function of the vertical mass column density $\chi$.}
         \label{fig:ionization_rate}
    \end{figure}

There are several sources of primary ionization such as 
galactic cosmic rays, UV and X-rays from the host star,
heat caused by stellar radiation or disk viscosity, and the decay
of short-lived radionuclides. Here we take into account cosmic 
rays, X-rays, and radionuclides as ionization sources. 
If we denote the radius of a disk as $r$ and the height from the 
disk mid-plane as $z$, 
the ionization rate can be written as
\begin{eqnarray}
    \zeta(r,z) & = & \zeta_{\rm C} + \zeta_{\rm X}
                + \zeta_{\rm R},
\label{231}
\end{eqnarray}
where $\zeta_{\rm C}$, $\zeta_{\rm X}$, and $\zeta_{\rm R}$
are the ionization rates of cosmic rays, X-rays, and 
radionuclides respectively. $\zeta_{\rm C}$ can be calculated
from the following equation \citep{ume81}:
\begin{eqnarray}
    \zeta_{\rm C} & = & \frac{\zeta_{\rm CR}}{2}
    \left\{ 
        \exp \left[-\frac{\chi(r,z)}{\chi_{\rm CR}}\right]
    \right.
         \left.+ \exp\left[ - \frac{\Sigma(r) 
            - \chi(r,z)}{\chi_{\rm CR}} \right]     
    \right\},
\label{232}
\end{eqnarray}
where $\zeta_{\rm CR} = 1.0\times 10^{-17}$ s$^{-1}$ is the cosmic ray 
ionization rate in interstellar space. Cosmic rays may be blown out 
by stellar winds and the ionization rate may be lower
by several orders of magnitude depending on stellar activity \citep{cle13}.
$\chi_{\rm CR}=96\ {\rm g\ cm^{-2}}$ is the attenuating length of cosmic rays, 
and $\Sigma(r)$ is the surface density of a disk at radius $r$, and
\begin{eqnarray}
    \chi(r,z) = \int_{z}^{\infty}\rho_{\rm g}(r,z)\ dz
\label{233}
\end{eqnarray}
is the column density from $z$ to the outside of the disk.
The rate of X-ray ionization is
\begin{equation}
    \zeta_{\rm X} = \zeta_{\rm XR}\left(\frac{r_{\ast}}{1{\rm AU}}\right)^{-2}
                    \left( \frac{L_{\rm XR}}{2\times10^{30}{\rm erg\ s^{-1}}} \right)
                    \left\{ \exp \left[-\frac{\chi(r,z)}{\chi_{\rm XR}}\right]
                    + \exp\left[ -\frac{\Sigma(r) -\chi(r,z)}{\chi_{\rm XR}}\right]
                    \right\},
    \label{234}
\end{equation}
where $r_{\ast}$ is the distance from the host star,
$L_{\rm XR} =  2\times 10^{30}$ erg s$^{-1}$ is the X-ray
luminosity, and 
$\zeta_{\rm XR} = 2.6\times10^{-15}$ s$^{-1}$ and
$\chi_{\rm XR} =  8.0$ g cm$^{-2}$ are fitting parameters
\citep{ige99, tur08}. Equation (\ref{234}) only takes into 
account X-rays scattered by diffuse gas well above 
the mid-plane and neglects direct X-ray irradiation 
\citep[see][]{tur08}. We use this formula since 
a protoplanetary disk likely blocks direct irradiation and 
prevents it from reaching a geometrically much thinner 
circumplanetary disk. A new study on X-ray ionization rates 
done by \citet{erc13} confirmed the calculation of \citet{ige99}
and also calculated ionization rates using parameters 
based on current observations.
The ionization rate of the decay of radionuclides is 
$\zeta_{\rm R}=7.6\times10^{-19}f_{\rm g}\ {\rm s}^{-1}$
where $f_{\rm g}$ is the depletion factor of dust grains 
from interstellar abundance \citep{ume09}. A plot of 
each ionization source is provided in Fig. \ref{fig:ionization_rate}.

If the density is high around the border of an MRI-active region 
and a dead zone, the main ionization source is cosmic rays and 
other sources are less effective at the orbit of gas giants. 
Large uncertainties exist about how many radionuclides 
are in circumplanetary disks and also how the effectiveness 
of X-ray ionization depends on the geometry of  
circumplanetary and protoplanetary disks.
However, here we include X-rays and radionuclides as well as 
cosmic rays to obtain a maximum estimate of ionization degree. 

\section{Disk model} \label{sec:CPDmodel}
\subsection{Surface density of circumplanetary disks}
\label{sec:surface density}

We solve a diffusion equation for a disk with mass infall from outside
the disk assuming accretion stress $\alpha$ \citep{sha73}
and determine the surface density of the disk. 
The diffusion equation of surface density $\Sigma$ in 
a Keplerian disk with a source term $f$ is
\begin{eqnarray}
    \frac{\partial \Sigma}{\partial t}
    =  \frac{1}{r}\frac{\partial }{\partial r}
    \left[ 3r^{1/2}\frac{\partial}{\partial r}
            \left( r^{1/2}\nu\Sigma \right)\right]
            + f,
    \label{311}
\end{eqnarray}
where $r$ is radius and $\nu$ is the kinematic 
viscosity coefficient. 
For $f$, we use the result of \citet{tan12}. They measured the physical
properties of infalling gas just before it falls onto a circumplanetary 
disk. Since the infall is supersonic, its properties do not depend on
the structure of the circumplanetary disk located at the downstream and thus 
the physical properties are less uncertain. 
They found that the mass flux and specific angular momentum of infalling gas  
are proportional to $r^0$ and $r^1$, respectively.
The angular momentum at the radius where gas falls onto the 
circumplanetary disk is smaller than that of Keplerian rotation 
and the radial dependence of angular momentum of infalling gas 
is larger than the Keplerian profile: gas will move inwards until it
rotates at the Keplerian velocity if it conserves specific angular momentum,
and the mass distribution becomes centrally concentrated.
As a result, the effective mass flux can be approximated as  $f \propto r^{-1}$.
We assume that a central planet is located at an orbit of $5.2$AU 
in a protoplanetary disk around a solar mass star.
If we adopt the minimum mass solar nebula model \citep{hay81},
the surface density and sound speed of the protoplanetary disk 
at 5.2AU are 
$\Sigma_{\rm P} = 143\ {\rm g\ cm^{-2}}$ and
$c_{\rm sP} = 6.58\times 10^4\ {\rm cm\ s^{-1}}$.
We assume the central planet has a mass of 0.4 Jupiter masses. 
Using these values, we obtain a mass infall rate of
\begin{eqnarray}
    f = 1.3\times10^{-3}\epsilon
        \left( \frac{\Sigma_{\rm p}}{143\ {\rm g\ cm^{-2}}} \right)
        \left( \frac{r}{R_{\rm J}} \right)^{-1}
        \ {\rm g\ cm^{-2}\ s^{-1}},
    \label{522}
\end{eqnarray}
where $\epsilon$ is a depletion factor of protoplanetary disk gas, and 
$R_{\rm J}$ is the Jupiter radius.
We use this formula only within $r=20 R_{\rm J}$ 
and set $f=0$ at larger radii, because the power law index of the mass infall 
rate drops outside $\sim20 R_{\rm J}$ \citep{tan12}. 
The parameter $\epsilon$ represents the situation when a (proto-) planet 
grows to some extent, and a gap opens in the disk. 
Since the viscous timescale 
of a circumplanetary disk is sufficiently smaller than that of 
a protoplanetary disk, we can treat $\epsilon$ as a constant.
Smaller values of $\epsilon$ represent later times, and 
$\epsilon=1$ corresponds to the onset of accretion.
We employ the standard $\alpha$ prescription,
\begin{eqnarray}
    \nu = \alpha c_{\rm s}H = \alpha\frac{c_{\rm s}^2}{\Omega_{\rm K}}.
    \label{524}
\end{eqnarray}
%
We choose $\nu$ such that $\alpha=0.05$. 

We solve Equation (\ref{311}) numerically with
the initial condition $\Sigma(t=0)=0$, time step
$\Delta t=1.0\times10^3\ {\rm s}$,
and cell width $\Delta r=0.85 R_{\rm J}$.
The calculation range is $0.85 R_{\rm J}\leq 
r \leq 210 R_{\rm J}$.
The boundary conditions are that the torque vanishes
at the center and at the outer boundary.

We assume that the disk is vertically hydrostatic and 
use the gas density profile 
\begin{eqnarray}
    \rho_{\rm g}(r,z) \equiv \frac{\Sigma}
        {\sqrt{\mathstrut 2\pi}H}
        \exp\left( -\frac{z^2}{2 H^2} \right),
\label{33}
\end{eqnarray}
and use $T=$123K as the temperature of the disk gas.

\begin{figure}[htpb]
    \epsscale{0.80}
    \plotone{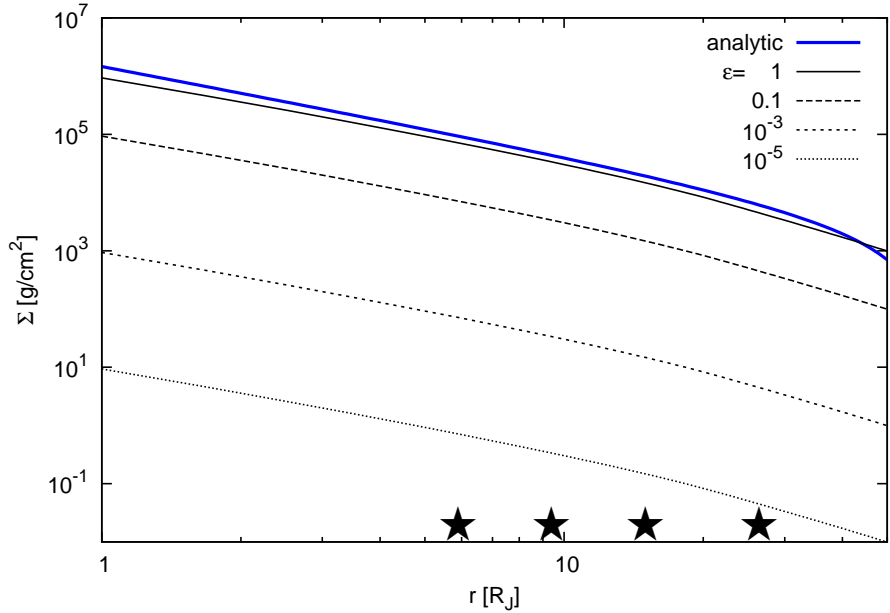}
         \caption{Surface densities of circumplanetary disks
         with $\alpha=0.05$. Each line shows a different 
         value of $\epsilon$, the depletion factor of 
         protoplanetary disk gas. The blue thick line is an analytic
         solution (see \ref{sec:CPDmodel2}). Stars denote the orbit of 
         Galilean satellites around Jupiter.}
         \label{fig:alpha=0.05}
    \end{figure}

In Fig. \ref{fig:alpha=0.05} we plot the surface density 
of steady states with 
$\epsilon=1,\ 0.1,\ 10^{-3}$, and $10^{-5}$.
Note that the critical surface density to be gravitationally unstable is 
several orders of magnitude larger than the case of $\epsilon=1$.
%
\subsection{Analytic solution for surface density} 
\label{sec:CPDmodel2}
We can analytically solve Equation (\ref{311})
using Equation (\ref{524}) when $\alpha$ is a constant value
and the disk is steady and isothermal \citep[cf. Appendix of][]{can02}. The source term $f$ is proportional
to $r^{-1}$ so we write $f\equiv A/r$, where A is a constant. If we write 
$\nu=\nu_0\left( r/r_0 \right)^{\frac{3}{2}}$, where 
$\nu_0$ is the kinematic viscosity coefficient for $r=r_0$,
the solution of 
\begin{eqnarray}
   0 = \frac{1}{r}\frac{\partial }{\partial r}
    \left[ 3r^{\frac{1}{2}}\frac{\partial}{\partial r}
            \left( r^{\frac{1}{2}}\nu\Sigma \right)\right]
            + \frac{A}{r},
    \label{5281}
\end{eqnarray}
is 
\begin{eqnarray}
    \Sigma = \frac{r_0^{\frac{3}{2}}}{\nu_0}\left( -\frac{2}{9}
                Ar^{-\frac{1}{2}}
            + C_1r^{-\frac{3}{2}}
            + C_2r^{-2} \right),
    \label{529}
\end{eqnarray}
where $C_1$ and $C_2$ are constants.
$C_2$ should be zero since the torque $r^{1/2}\nu\Sigma$ 
vanishes at $r=0$.
The mass accretion rate of a steady state at the inner boundary,
$\dot M_{\rm p}$, is
\begin{eqnarray}
    {\dot M_{\rm p}} 
    &\simeq& -2\pi r_{\rm in} \Sigma v_{\rm r}\nonumber\\
    &=& 6\pi r_{\rm in}^{\frac{1}{2}}\frac{\partial}{\partial r}
          \left( r_{\rm in}^{\frac{1}{2}} \nu \Sigma\right)\nonumber\\
    &=& 3\pi C_1,
    \label{5210}
\end{eqnarray}
where $r_{\rm in}$ is the radius of the inner boundary and 
$v_{\rm r}$ is the radial velocity of gas. We consider the 
second term on the right hand side of Equation (\ref{529}) to be dominant. 
%
Since the inner edge of a disk is far smaller than the outer edge of
the region with infall, $r_{\rm b}$, the total infall rate 
onto the circumplanetary disk $\dot{M}_{\rm s}$
can be approximated as
\begin{eqnarray}
    \dot M_{\rm s} &=& 2\pi\int^{r_{\rm b}}_{r_{\rm in}} f r\ dr\nonumber\\
                   &\simeq& 2\pi Ar_{\rm b}.
    \label{5211}
\end{eqnarray}
In a steady state, the mass falling onto the central planet should be equal 
to the inflow from the surrounding protoplanetary disk, thus 
$\dot M_{\rm p} = \dot M_{\rm s}$ and $C_1$ can be derived as 
\begin{eqnarray}
    C_1 = \frac{2}{3}Ar_{\rm b}.
    \label{5212}
\end{eqnarray}
In this way, we obtain the analytic solution for the surface density:
\begin{eqnarray}
    \Sigma = \frac{A}{\nu_0}
             r_0^{\frac{3}{2}}
            \left( -\frac{2}{9}r^{-\frac{1}{2}} + 
                   \frac{2}{3}r_{\rm b}r^{-\frac{3}{2}} \right).
    \label{5213}
\end{eqnarray}
This solution for $\epsilon=1$ and $r_{\rm b}=20R_{\rm J}$
is plotted in Fig. \ref{fig:alpha=0.05}.
We consider only the mass accreted onto the central planet and 
neglect the mass extending outwards, thus the analytic solution 
is slightly larger than the numerical solution.
About 10\% of the infalling mass flux exits the disk outwards.
\section{Ionization degree and MRI-activity in circumplanetary disks}
\label{sec:MRIinCPD}
We calculate the ionization degree in circumplanetary disks for 
the surface densities obtained in Section \ref{sec:CPDmodel}.
The parameters employed in our calculations are shown in 
Table \ref{tbl-1}.
Note that cases with larger $f_{\rm dg}$
or smaller $a$ than the ranges shown result in smaller MRI-active regions
and the results with smaller $f_{\rm dg}$ or larger $a$ approach those
of the dust-free calculations. Since the magnetic field strength is uncertain, 
we choose optimistic values of $\beta_0$.
If we focus only on the MRI, there are no heating sources if the disk is MRI-dead. 
The radial profiles of temperature structure do not dramatically affect
MRI-activity as long as the disk is not hot enough for thermal ionization
to be effective. Thus, we assume isothermality in our calculations of 
ionization degree to be consistent with the earlier sections of this paper.
\clearpage

\begin{figure}[t]
    \epsscale{1.1}
    \plottwo{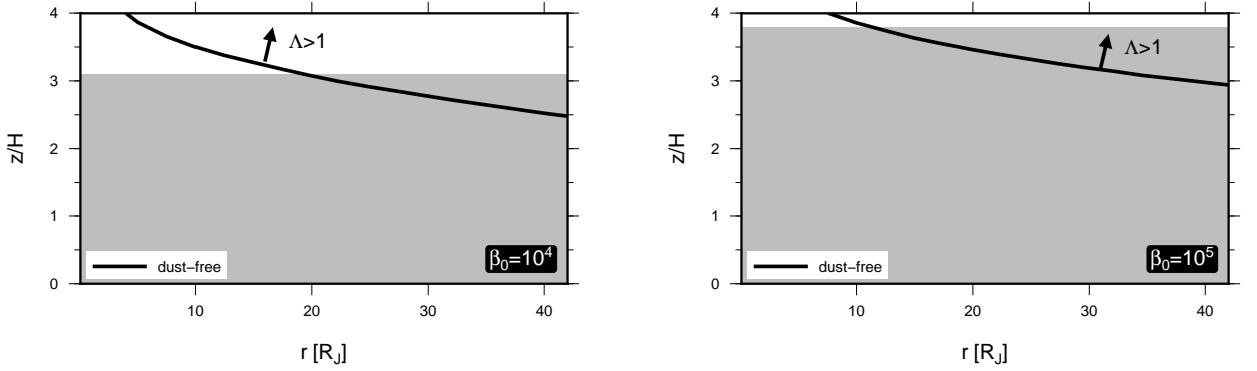}{tcpd_isoX-nu-3dep1beta5-new.eps}
         \caption{Borders of MRI-active/inactive regions.
         The horizontal axis is radius normalized by Jupiter radius,
         and the vertical axis is the vertical extent of the disk 
         normalized by disk scale height.
         The solid lines show $\Lambda=1$ for the dust-free case, and
         the region above the lines are $\Lambda>1$. Shaded and hatched 
         areas are the regions $\beta_{\rm z}>8\pi^2$ and 
         $\beta_{\rm z}>2000$, respectively.
         The left panel is the case with $\beta_0=10^4$ and 
         the right panel is the case with $\beta_0=10^5$.
         These are cases in which the infall rate is not 
         decreased ($\epsilon=1$).
         We plot only the results of dust-free cases. If there
         are dust grains, the ionization degree is smaller and 
         the line $\Lambda=1$ is higher. Gas from the protoplanetary disk
         infalls onto the region $r<20\ R_{\rm J}$.}
         \label{fig:f_dp=1}
    \end{figure}

\begin{deluxetable}{ll}
\tabletypesize{\scriptsize}
\tablecaption{Calculations parameters\label{tbl-1}}
\tablewidth{0pt}
\tablehead{
\colhead{parameter} & \colhead{value}
}
\startdata 
    gas depletion factor $\epsilon$ & 1, 10$^{-3}$, 10$^{-5}$ \\
    dust-to-gas ratio $f_{\rm dg}$ & 0 (dust-free), 0.01 \\ 
    dust radius $a [\mu {\rm m}]$ & 0.1, 10 \\
    mid-plane value of plasma beta ($z$ component) 
        $\beta_{\rm 0}$ & 10$^4,$ 10$^5$ 
\enddata 
\end{deluxetable}
Fig. \ref{fig:f_dp=1} shows the results for the case of 
$\alpha=0.05$, $\epsilon=1$, and $f_{\rm dg}=0$.
The region above the solid line $\Lambda=1$ and inside the gray shaded area   
($\beta_{\rm z}>8\pi^2$) is unstable to MRI as dictated by the conditions 
discussed in Section \ref{sec:2-1}. 
The hatched region illustrates $\beta_{\rm z}>2000$, the criterion for turbulence
to be well developed (refer to Equation \ref{2000}).
Thus, a region of well-developed MRI turbulence is above $\Lambda=1$
and within the hatched region. For the parameters of Fig. \ref{fig:f_dp=1}, 
there is no region that has well-developed MRI turbulence. 
This means that under these settings, 
the MRI cannot generate the accretion stress, $\alpha=0.05$,
which we have assumed in the calculations of surface density.
A calculation with smaller $\alpha$ results in a smaller 
region with $\Lambda>1$ because the surface density is 
larger and the ionization degree is lower.
Consequently, we cannot find a self-consistent solution for 
$\Sigma$ and $\alpha$ when $\epsilon=1$.

\begin{figure}[t]
    \epsscale{1.1}
    \plottwo{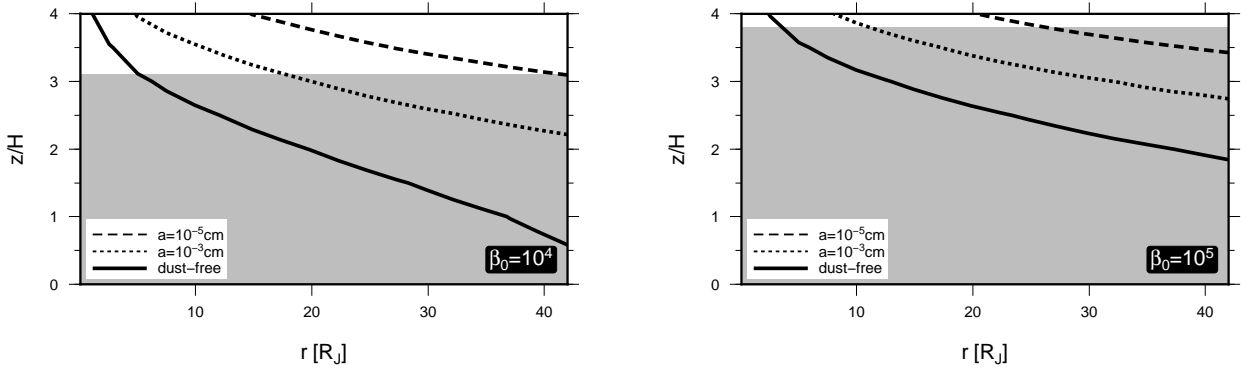}{tcpd_isoX-nu-3dep-3beta5-new.eps}
         \caption{Same as Figure \ref{fig:f_dp=1} but for the case of 
         $\epsilon=10^{-3}$. The radius of dust grains, $a$, 
         used in these calculations is indicated in the figure.}
         \label{fig:f_dp=10^-3}
    \end{figure}

\begin{figure}[t]
    \epsscale{1.1}
    \plottwo{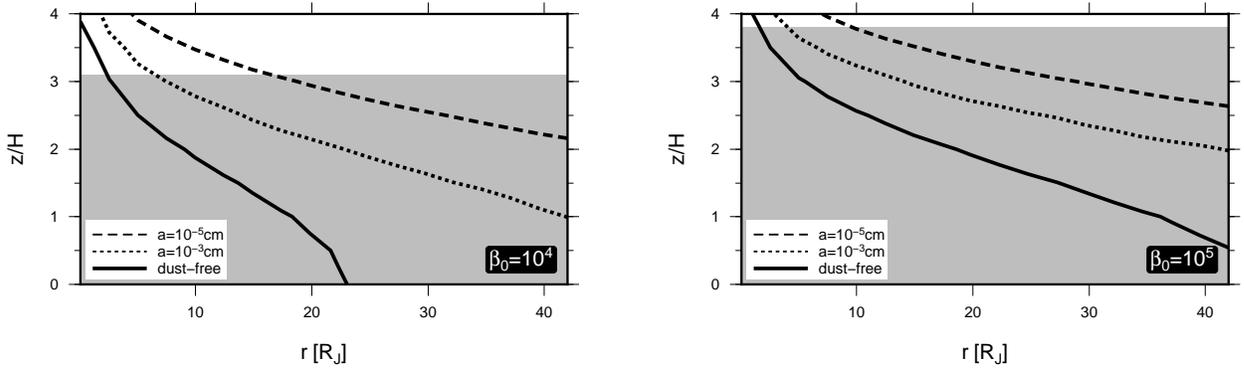}{tcpd_isoX-nu-3dep-5beta5-new.eps}
    \caption{Same as Fig. \ref{fig:f_dp=10^-3} but for the case of 
    $\epsilon=10^{-5}$.}
         \label{fig:f_dp=10^-5}
\end{figure}

The results for $\epsilon=10^{-3}$ are shown in 
Fig. \ref{fig:f_dp=10^-3}, which corresponds to the case of 
gap opening or global disk dispersal.
Without dust grains, active layers with well-developed turbulence 
appear at large radii, but with dust grains, 
such layers do not exist in satellite-forming regions.
Fig. \ref{fig:f_dp=10^-5} shows the results for $\epsilon=10^{-5}$. 
The MRI-active layers become thicker but the situation 
does not change dramatically. It is difficult to sustain well-developed 
MRI turbulence in circumplanetary disks with dust grains, especially in 
areas experiencing gas infall.
%

When the surface density is smaller, the ionization degree is larger
and the line $\Lambda=1$ is lower. We estimate the minimum
surface density needed to form satellites to find the lower 
limit of the line $\Lambda=1$. 
The timescale to form a satellite of radius 
$R_{\rm s}$ and density $\rho_{\rm s}$ 
orbiting at $r$ is 
\begin{eqnarray}
    \tau_{\rm s}&\approx&\frac{1}{\Omega}
    \frac{\rho_{\rm s}R_{\rm s}}{\Sigma_{\rm sol}F_{\rm g}}\nonumber\\ 
    &\sim& 8{\rm yr}\left( \frac{R_{\rm s}}{2500\ {\rm km}} \right)
    \left( \frac{\rho_{\rm s}}{2\ {\rm g\ cm^{-3}}} \right)
    \left( \frac{F_{\rm g}}{100} \right)^{-1}\nonumber\\
    &\ &\times
    \left( \frac{\Sigma_{\rm sol}}
       {3\times10^3\ {\rm g\ cm^{-2}}} \right)^{-1}
    \left( \frac{r}{15R_{J}} \right)^{3/2},
    \label{43}
\end{eqnarray}
where $\Sigma_{\rm sol}$ is the surface density of solids and 
$F_{\rm g}\equiv1+(v_{\rm esc}/v_{\infty})^2$ is 
the gravitational focusing factor for colliding objects 
with relative velocity at infinity $v_{\infty}$ and 
 mutual escape velocity $v_{\rm esc}$ \citep{lis93b, war96, can02}.
To form a satellite whose radius, density, 
and orbit are similar to Ganymede's within the lifetime of 
the disk, $\tau_{\rm s}\sim 10^7$ yr, the required surface 
density of solids is
\begin{equation}
    \Sigma_{\rm sol}\sim 10^{-3}\ {\rm g\ cm^{-2}}
     \left( \frac{r}{15R_{J}} \right)^{3/2}
     \left( \frac{R_{\rm s}}{2500\ {\rm km}} \right)
    \left( \frac{\rho_{\rm s}}{2\ {\rm g\ cm^{-3}}} \right)
    \left( \frac{F_{\rm g}}{100} \right)^{-1}.
    \label{44}
\end{equation}
The value of $F_{\rm g}$ depends on the size of the proto-satellite.
Since we want to know the minimum value of $\Sigma_{\rm sol}$, 
we choose the maximum value of the gravitational focusing factor, 
$F_{\rm g}\sim100$, when $v_{\infty}\simeq
\left( M_{\rm s}/3M_{\rm p} \right)^{1/3}v_{\rm K}$ ($v_{\rm K}$
is the Keplerian velocity around the planet at the proto-satellite orbit).
If we assume the dust-to-gas mass ratio does not depend on the height and 
take $f_{\rm dg}=10^{-2}$, the surface density of gas
should be larger than $\Sigma\sim 0.1\ {\rm g\ cm}^{-2}$.
According to this estimation, and Fig. \ref{fig:alpha=0.05}, 
it seems quite difficult to form satellites 
with $\epsilon=10^{-5}$ or smaller. 
Therefore, we do not consider even smaller infall rates.

\begin{figure}[htpb]
    \epsscale{1.00}
    \plottwo{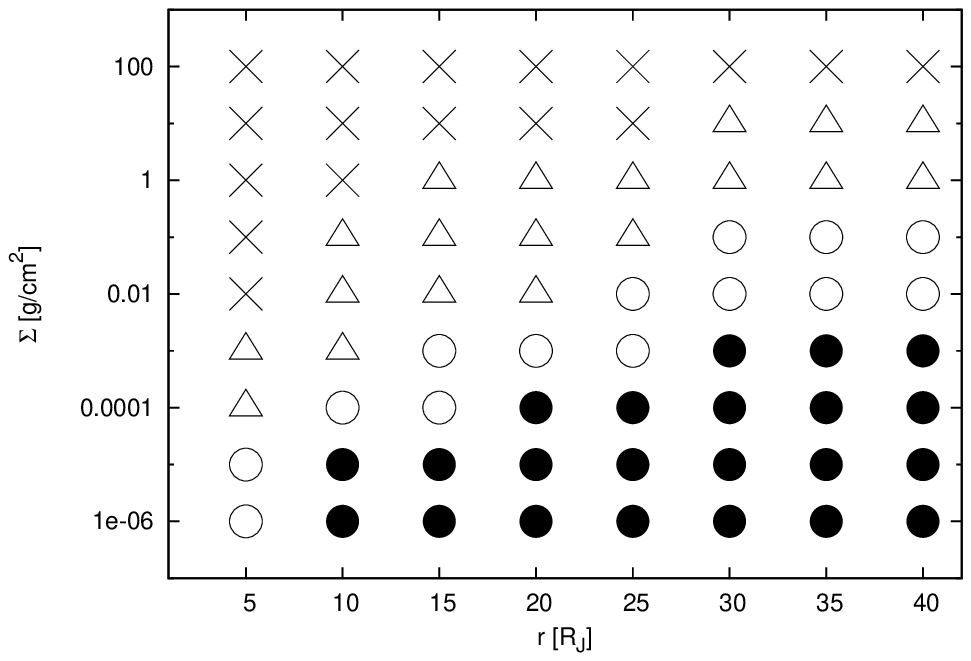}{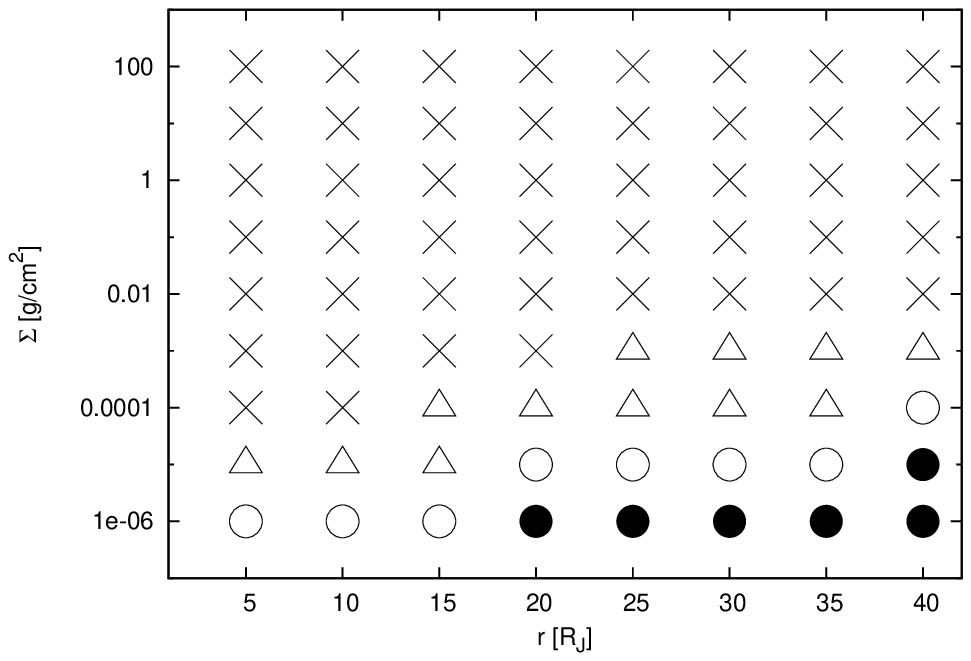}
         \caption{Surface densities of MRI-active at each 
         radius. $\bullet$ represents the surface densities at the 
         given radius which have well-developed MRI turbulence 
         at the entire height, $\circ$ represents that at only above $0.5H$, 
         and $\triangle$ represents that at only above $2H$. 
         Since we choose $\beta_0 = 10^5$, $\beta_{\rm z}<2000$
         above $2.8H$ and well-developed turbulence cannot be sustained
         in such region. We plot $\times$ for the surface densities
         in which we cannot find regions of $\Lambda>1$ below $2.5H$.
         The left panel is the result of a dust free calculation and
         the right panel is that of a case with dust grains 
         of $a=0.1\mu$m. 
         With dust grains, the MRI-active region is smaller because 
         of the lower ionization degree. }
         \label{fig:MRI}
    \end{figure}

Next, we investigate the surface density at each radius 
that can sustain well-developed MRI turbulence. The results are summarized in
Fig. \ref{fig:MRI}. The surface densities which can sustain well-developed
turbulence for $z>2H$, $z>0.5H$, and for the entire height at each radius are shown.
The mid-plane plasma beta considered here is $\beta_0=10^5$.
If we choose larger $\beta_0$, the line $\Lambda=1$ shifts higher 
which means the MRI-active region becomes smaller.
On the other hand, if we choose smaller $\beta_0$, 
the region with $\beta_{\rm z}>2000$ is smaller,
and having large MRI-active regions becomes difficult. 
%
\section{Discussion}
\label{sec:5}
We find that when accounting for cosmic rays, X-rays, and radionuclides,
circumplanetary disks are not likely to sustain well-developed MRI turbulence.
In contrast to protoplanetary disks, the volume ratio of MRI-active 
regions to dead zones for circumplanetary disks is very small.
This is because the typical length scale of circumplanetary disks
is smaller by several orders of magnitude than that of 
protoplanetary disks. This makes the timescale of magnetic diffusion
smaller. According to Equation (\ref{411}), 
even if $v_{{\rm A}z}$ and $\eta$ are the same, the Elsasser 
number tends to be smaller in circumplanetary disks.
For example, the typical ionization degree at 5AU in a protoplanetary disk
and that at 15$R_{\rm J}$ in a circumplanetary disk are both $\sim 10^{-10}$,
but the Keplerian frequency of a protoplanetary disk at 5AU is $\sim 10^{-8}$ 
s$^{-1}$ and that of a circumplanetary disk at 15$R_{\rm J}$ 
is $\sim 10^{-5}$ s$^{-1}$.  
Therefore, it is very difficult for MRI turbulence to be well developed
in circumplanetary disks unless the surface density is very small, as 
we show in Fig. \ref{fig:MRI}. 

Recently, \citet*{tur13} have investigated the possibility of
the MRI in various models of circumplanetary disks from the literature.
They choose the mid-plane value of plasma beta to be $10^3$. 
Their condition to have the MRI is that the magnetic pressure is smaller 
than the gas pressure, which is satisfied below 3.7 scale heights.
They concluded that there are active layers at the disk surface. 
Our results are consistent with theirs 
when we choose the condition that $\beta_{\rm z}>8\pi^2$ to sustain
the MRI. For example, Figure 3 of \citet{tur13} 
and Fig. \ref{fig:f_dp=1} of this paper show similar models 
of surface density, and both have surface active layers.
Note that even if the MRI can be sustained at the disk
surface, it does not necessarily mean that there is well-developed turbulence.
When MRI turbulence is well developed, the magnetically dominated 
atmosphere encroaches on a lower altitude and a region of well-developed
turbulence becomes smaller \citep{oku13a}. The main difference between 
\citet{tur13} and our work is that we consider the criterion for turbulence
to be well developed as well as that of just having MRI.

If there are no other mechanisms to give rise
to viscosity and disks are not massive, 
the gas piles up in circumplanetary disks until the disks 
become massive enough to be gravitationally unstable. It is 
possible to promote gas accretion by GI,
but it may not reduce the disk surface density much below 
the critical value for GI. 
Therefore the surface density is expected to remain large.

As we mentioned, if only GI can generate gas accretion, a massive and 
static disk will remain even after infall from the protoplanetary
disk terminates. 
This suggests that the lifetime of circumplanetary disks may be 
longer than that of protoplanetary disks. Thus, perhaps we are 
more likely to be able to observe circumplanetary disks than previously 
thought \cite[e.g.,][]{mam12}, and satellite formation may occur over a long timescale.
On the other hand, it remains important to consider other 
mechanisms for angular momentum transport.
A possible mechanism is spiral density waves caused by a non-axisymmetric 
potential \citep[e.g.,][]{mac10, mar11b, riv12, szu13}, 
but this must be investigated in more detail. 

If there is viscous heating in such a massive disk, 
it will easily heat up, and thermal ionization may drive the MRI
\citep{lub12}.
Even if GI can drive turbulence, it does not necessarily mean
the turbulence can generate heat in situ. The question of where the 
energy dissipates remains open \citep{bal99, god01, mut10}. 
We should be careful in treating the heating by gravitational turbulence.
Further study of energy dissipation by GI is necessary. 

If thermal ionization triggers the MRI, disks would be less massive 
because of a high accretion rate.
Suppose that if at each annulus of a disk, gravitational energy 
is converted into thermal energy and radiates as a black body from the disk
surface, the effective temperature is  
\begin{eqnarray}
    T_{\rm eff} = \left( \frac{3GM_{\rm p}\dot M}
                    {8\pi\sigma_{\rm SB} r^3} \right)^{1/4},
    \label{5-1}
\end{eqnarray}
where $M_{\rm p}$ is the planet mass, $\dot M$ is the mass accretion rate,
and $\sigma_{\rm SB}$ is the Stefan-Boltzmann constant. 
We can estimate the mid-plane temperature, $T_{\rm c}$, from the 
approximation 
$T_{\rm c} \simeq \tau^{1/4} T_{\rm eff}\ (\tau \gg 1)$,
where $\tau$ is optical depth, given by 
$\tau\sim\kappa\Sigma$ where $\kappa$ is opacity.  
Here we use  $\dot M = 3\pi\nu\Sigma$. 
Then, the mid-plane temperature can be written as 
\begin{eqnarray}
    T_{\rm c} = 1.0\times10^3\left( \frac{\kappa}
                    {5\ {\rm cm^2\ g^{-1}}} \right)^{1/5}
                \left( \frac{\alpha}{10^{-2}} \right)^{-1/5}
                \left( \frac{M_{\rm p}}{M_{\rm J}} \right)^{3/10}\nonumber\\
                \times \left( \frac{\dot M}
                {3.1\times10^{-7}M_{\rm J}\ {\rm yr}^{-1}} \right)^{2/5}
                \left( \frac{r}{10R_{\rm J}} \right)^{-9/10}\ {\rm K},
    \label{5-2}
\end{eqnarray}
where ${M_{\rm J}}$ is the Jupiter mass.
When the mid-plane temperature exceeds about 
1000K, gas in the disk will be sufficiently ionized to have the MRI.
The inner disk may be hot enough to have thermal ionization especially 
during the early phases of gas giant formation; however, the outer disk seems 
to remain cool. More detailed calculations of the mid-plane temperature 
of circumplanetary disks have been done by \citet{kei14}.

Another possibly important mechanism for satellite formation is 
the capture of planetesimals when they cross circumplanetary 
disks \citep{fuj13} \citep[see][for the gas-poor case]{est06}. 
Since our results suggest a large surface 
density, capture is expected to be effective. Non-axisymmetry 
in the density structure caused by these 
proto-satellites may play a role in angular momentum transport and
it may be interesting to analyze that effect.
\section{Summary}
\label{sec:6}
We estimated the size of regions that can sustain magnetic
turbulence in circumplanetary disks. We calculated the ionization
degree in disks accounting for galactic cosmic rays, X-rays from
the host star of the surrounding protoplanetary disk, and the decay of 
short-lived radionuclides as ionization sources, and evaluated 
the MRI activity.  We adopted the $\alpha$ model and 
solved the diffusion equation of a disk with infalling mass flux 
from a protoplanetary disk, obtained by \citet{tan12}.
Even by varying parameters such as gas infall rate, magnetic field, 
dust-to-gas mass ratio, and radius of dust grains over a wide range, it was 
difficult to find a sufficiently sized MRI-active region that can sustain
well-developed turbulence, a region where both the Elsasser number is larger 
than unity and the magnetic pressure is sufficiently smaller than the 
gas pressure.  We found that the surface density that can sustain 
well-developed MRI turbulence is 
$\Sigma\sim$0.001-0.01 g cm$^{-2}$, even without dust grains,
for a typical satellite-forming region. 
Note that we have performed the calculations for a very optimistic 
set of assumptions  for activation of the MRI. If metals are 
frozen out onto dust grains or cosmic rays are shielded by 
stellar activities, situations become much severer to sustain MRI.

If there are MRI-active regions, we can estimate the accretion 
stress with the empirical formula of \citet{oku11, oku13a}. 
However, we find that the MRI is unlikely to be well developed 
in circumplanetary disks with cosmic rays, X-rays, and radionuclides; 
even if the MRI can be initiated, the active turbulence 
cannot be sustained. 
As long as the MRI is the only mechanism of gas accretion 
in less massive disks, our results suggest that disk surface 
density increases until it becomes gravitationally unstable.
If this is the case, the picture of satellite formation should be changed.
In order to examine this suggestion, we have to
investigate other possible mechanisms such as spiral density waves, 
baroclinic instability, or global magnetic braking.  

In this paper, we have used the analysis of \citet{tan12} for the  
formation phases of circumplanetary disks. However, the long-term  
evolution of the gas infall rate from a protoplanetary disk to 
a circumplanetary disk is not yet well understood. 
To investigate this further, it is necessary to know the evolution of 
protoplanetary disks, such as the gap opening timescale and its effect
on density structure. We need to study how a change in the density
of protoplanetary disks affects the infall rate onto circumplanetary 
disks. 

\acknowledgments
                    
We thank Sanemichi Z. Takahashi and Hiroshi Kobayashi for 
fruitful discussions. 
We are grateful to the anonymous referees for useful comments 
that improved our manuscript.
We appreciate Jennifer M. Stone's help in improving our English,
and Takeru K. Suzuki's continuous encouragement. 
This work was supported by JSPS KAKENHI Grant Numbers 24$\cdot$4770,
and 25887023, and by MEXT KAKENHI Grant Numbers 23740326, 24103503,
23244027, and 23103005.


\end{document}